\documentclass[conference,a4paper]{IEEEtran}
\IEEEoverridecommandlockouts
\def\BibTeX{{\rm B\kern-.05em{\sc i\kern-.025em b}\kern-.08em
    T\kern-.1667em\lower.7ex\hbox{E}\kern-.125emX}}

\usepackage{amsmath,algorithm, algorithmic}
\usepackage{cite}
\usepackage{epsf,graphics,graphicx}
\usepackage{comment}
\usepackage{bm}
\usepackage{amssymb,amsthm}
\usepackage{amsmath}
\usepackage[top=.75in, left= 0.625in, right = 0.625in, bottom =1in]{geometry}
\newtheorem{assumption}{Assumption}

\newtheorem{identity}{Identity}

\newtheorem{Proposition}{Proposition}

\newtheorem*{remark*}{Remark}

\usepackage{comment}
\usepackage[textwidth=30mm]{todonotes}
\usepackage{soul}

\setstcolor{magenta}
\sethlcolor{lightgray}

\def\vec#1{{\bf #1}}
\def\Tr#1{\mathrm{Tr}\left\{ #1 \right\} }
\def\outer{\otimes}

\DeclareMathOperator{\ex}{\mathbb{E}}

\newcommand{\ZZ}{\mathbb{Z}}

\newcommand{\CC}{\mathbb{C}}

\def\bPhi{\mbox{\boldmath$\Phi$}}
\def\bSigma{\mbox{\boldmath$\Sigma$}}

\def\bA{{\bf A}}
\def\bB{{\bf B}}

\def\bG{{\bf G}}

\def\bI{{\bf I}}
\def\bQ{{\bf Q}}
\def\bR{{\bf R}}
\def\bS{{\bf S}}
\def\bT{{\bf T}}

\def\bV{{\bf V}}
\def\bW{{\bf W}}

\def\bY{{\bf Y}}
\def\bZ{{\bf Z}}

\def\nt{{N_t}}
\def\nr{{N_r}}
\def\ns{{N_s}}

\def\bRcal{\mathbfcal{R}}
\def\bTcal{\mathbfcal{T}}

\DeclareMathAlphabet\mathbfcal{OMS}{cmsy}{b}{n}

\def\be{{\bf e}}
\def\bk{{\bf k}}

\def\bq{{\bf q}}

\def\bs{{\bf s}}

\def\bu{{\bf u}}
\def\bv{{\bf v}}

\def\bx{{\bf x}}
\def\by{{\bf y}}
\def\bz{{\bf z}}

\begin{document}
\bstctlcite{BSTcontrol}

\title{Capacity Optimization using Reconfigurable Intelligent Surfaces: A Large System Approach}
\author{Aris L. Moustakas$^1$, George C. Alexandropoulos$^2$, and M\'{e}rouane~Debbah$^3$\\
$^1$Department of Physics, National and Kapodistrian University of Athens, Greece
\\$^2$Department of Informatics and Telecommunications, National and Kapodistrian University of Athens, Greece
\\$^3$Technology Innovation Institute, Abu Dhabi, United Arab Emirates
\\e-mails: arislm@phys.uoa.gr, alexandg@di.uoa.gr, merouane.debbah@tii.ae
\vspace{-0.45cm}
}

\maketitle

\begin{abstract}
Reconfigurable Intelligent Surfaces (RISs), comprising large numbers of low-cost and passive metamaterials with tunable reflection properties, have been recently proposed as an enabler for programmable radio propagation environments. However, the role of the channel conditions near the RISs on their optimizability has not been analyzed adequately. In this paper, we present an asymptotic closed-form  expression for the mutual information of a multi-antenna transmitter-receiver pair in the presence of multiple RISs, in the large-antenna limit, using the random matrix and replica theories. Under mild assumptions, asymptotic expressions for the eigenvalues and the eigenvectors of the channel covariance matrices are derived. We find that, when the channel close to an RIS is correlated, for instance due to small angle spread, the communication link benefits significantly from the RIS optimization, resulting in gains that are surprisingly higher than the nearly uncorrelated case. Furthermore, when the desired reflection from the RIS departs significantly from geometrical optics, the surface can be optimized to provide robust communication links. Building on the properties of the eigenvectors of the covariance matrices, we are able to find the optimal response of the RISs in closed form, bypassing the need for brute-force optimization.
\end{abstract}

\begin{IEEEkeywords}
Reconfigurable intelligent surface, multipath, beamforming, capacity, MIMO, random matrix theory, replicas.
\end{IEEEkeywords}


\section{Introduction}\vspace{-0.1cm}
Future wireless networks are expected to transform to a unified communication and computing platform with embedded intelligence, enabling sixth Generation (6G) service requirements \cite{Samsung}. 
To accomplish this overarching goal, advances at various aspects of the network design are necessary, including wideband front-ends and smart wireless connectivity schemes \cite{Samsung}. Reconfigurable Intelligent Surfaces (RISs) \cite{huang2019holographic,DMA_2020} constitute a key wireless hardware technology for the recently conceived concept of ElectroMagnetic (EM) wave propagation control \cite{liaskos2018new,huang2019reconfigurable,RIS_Scattering,WavePropTCCN}, which is envisioned to offer manmade manipulation of the wireless environment. This low-cost green technology enables coating the various obstacles and objects of the environment with ultra-thin RISs, thus, transforming them into network entities with dynamically reconfigurable properties for wireless communications.

Over the last few years, metamaterials have emerged as a powerful technology with a broad range of applications, including wireless communications \cite{RIS_Scattering,WavePropTCCN}. They constitute artificial elements with physical properties that can be engineered to exhibit various desired characteristics \cite{smith2}. When deployed in planar structures (a.k.a. metasurfaces), their effective parameters
can be tailored to realize desired reflections of their impinging EM waves \cite{smith3}. RISs are essentially surfaces comprising many small reflecting elements, which may be tuned independently to manipulate the surface's reflection properties \cite{Hum2014}. Over the last few years, metamaterials have emerged as candidates of such reflecting elements, since their EM properties can be dynamically altered \cite{Science_2011}.

Analyzing the performance gains that RISs can offer in wireless communications has lately attracted research attention \cite{Jung2020,Nadeem2020,Selimis2021}, as a means to unveil their true potential for 6G networks. However, the current studies focus on scenarios with a single RIS, consider availability of the instantaneous channels, and overlook the role of the channel conditions near the RIS. In this paper, we use the ergodic Mutual Information (MI) as a performance metric to analyze the potential gains from multiple RISs for a multi-antenna transmitter-receiver pair. We derive an analytic expression for this quantity, valid in the limit of large array sizes, using random matrix theory and statistical physics tools. We optimize the MI subject to knowledge of the statistical properties of the channel, which is a more realistic RIS optimization strategy, due to its size and channel fluctuations. Using the knowledge of the asymptotic form of the eigenvectors of the channel covariance matrices, we are able to directly optimize the multiple RISs.

\textit{Notations:} We use bold-faced upper-case letters for matrices, e.g., $\vec X$ with its $(i,j)$-element denoted by $[\vec X]_{i,j}$, and bold-faced lower-case letters for column vectors, e.g., $\vec x$ with its $i$-element represented by $[\vec x]_i$. The superscripts $T$ and $\dagger$ indicate
transpose and Hermitian conjugate operations, $\Tr{\, \cdot\,  }$ is the trace operator, and $\vec I_n$ represents the $n$-dimensional identity matrix. The superscripts/subscripts $t$ and $r$ are used for quantities (e.g., channel matrices) referring to the Transmitter (TX) and Receiver (RX), respectively. $\mathbf{x}\sim{\cal CN}(\mathbf{0}_n,\mathbf{I}_n)$ denotes an $n$-element complex and circularly symmetric Gaussian vector with zero-mean elements and covariance matrix $\mathbf{I}_n$, while $\ex[\,\cdot\,]$ is the expectation operator. 

\section{System and Channel models}\label{sec:MIMO channel model}
\begin{figure}[!t]
	\centering
	\includegraphics[width=0.46\textwidth]{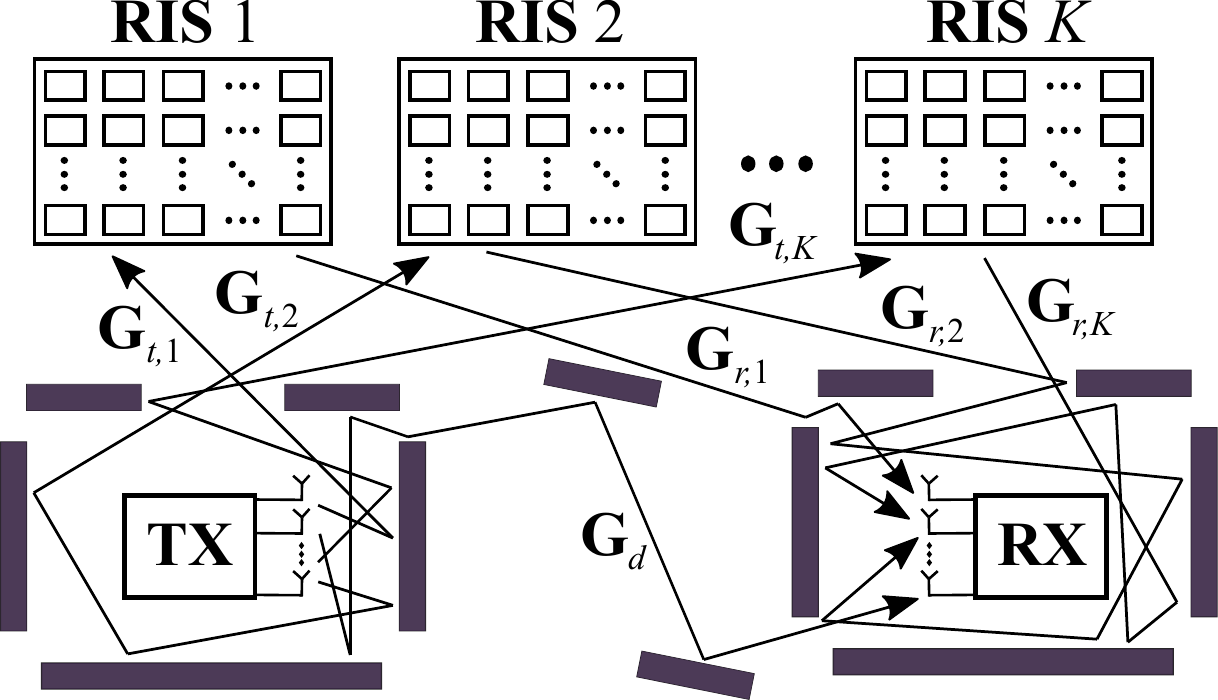}
	  \caption{The considered wireless communication system between an $\nt$-antenna TX and an $\nr$-antenna RX empowered by $K$ identical $\ns$-element RISs. The EM wave propagation environment may include obstacles (dark grey rectangulars) contributing to local scattering at the TX and RX.}\vspace{-0.4cm}
		\label{fig:system_model}
\end{figure}

\vspace{-0.1cm}
We consider the wireless communication system of Fig$.$~\ref{fig:system_model} between a TX equipped with $\nt$ antennas and an RX with an $\nr$-antenna array over a fading channel comprising their direct link as well as channels resulting from reflections from $K$ identical RISs, each consisting of $\ns$ tunable reflecting elements. We assume that all channels are known to RX via adequate estimation \cite{wang2020channel}, but not to TX. 
The baseband representation of the $\nr$-dimensional received signal vector at RX can be mathematically expressed as follows:
\begin{equation}\label{eq:basic_channel_eq}
  \by = \sqrt{\rho}\bG_{\rm tot}  \bx + \bz,
\end{equation}
where $\bx$ is the $\nt$-dimensional signal vector with covariance matrix $\bQ\triangleq \ex[\bx\bx^\dagger]$ normalized such that $\Tr{\bQ}=\nt$, and $\bz\sim{\cal CN}(\mathbf{0}_\nr,\mathbf{I}_\nr)$ is the noise vector.
The $\nr\times\nt$ channel gain matrix $\bG_{\rm tot}$ in our system model  \eqref{eq:basic_channel_eq} can be written as
\begin{align}\label{eq:Gtot}
\bG_{\rm tot} = \bG_d + \sum_{k=1}^K\sqrt{\gamma_k}\bG_{r,k}\bPhi_k\bG_{t,k},
\end{align}
where $\bG_{r,k}$ and $\bG_{t,k}$ with $k=1,2,\ldots,K$ represent the $\nr\times \ns$ and $\ns\times\nt$ channel matrices from the $k$-th RIS to RX and from the TX to the $k$-th RIS, respectively, while $\bG_d$ denotes the direct $\nr\times\nt$ channel matrix between the RX and TX, which is not reflected from any RIS. In addition, $\bPhi_k$ is the $\ns$-dimensional diagonal square matrix containing the tunable reflection coefficients at the $k$-th RIS in the main diagonal. The parameter $\rho$ in \eqref{eq:basic_channel_eq} represents the Signal-to-Noise Ratio (SNR) of the link, if only the direct channel matrix was present, while $\gamma_k$ is a measure of the relative additional SNR from each $k$-th RIS. Each $n$-th reflection coefficient ($n=1,2,\ldots,\ns$) of each $k$-th RIS is modeled as $[\bPhi_k]_{n,n}\triangleq e^{i\theta_{k,n}}$ \cite{huang2019reconfigurable}. We assume that all involved channel matrices in $\bG_{\rm tot}$ are complex Gaussian with the following Kronecker-product covariance matrices $\forall i,j=1,2,\ldots,\nr$, $\forall m,n=1,2,\ldots,\nt$, and $\forall a,b=1,2,\ldots,\ns$:
\begin{subequations}\label{eq:all_correlations}
\begin{align}
\label{eq:Gd_cov}
\ex\left[[\bG_d]_{i,m}[\bG_d]_{j,n}^*\right]&=\frac{1}{\nt}[\bR_d]_{i,j}, [\bT_d]_{m,n}
\\ \label{eq:Grk_cov}
\ex\left[[\bG_{r,k}]_{i,a}[\bG_{r,k}]_{j,b}^*\right]&=\frac{1}{\nt}[\bR_k]_{i,j} [\bS_{r,k}]_{a,b},
\\ \label{eq:Gtk_cov}
\ex\left[[\bG_{t,k}]_{a,m}[\bG_{t,k}]_{b,n}^*\right]&=\frac{1}{\nt}[\bS_{t,k}]_{a,b} [\bT_k]_{m,n}, 
\end{align}
\end{subequations}
where the received covariance matrices: $\bR_k$ of dimension $\nr\times \nr$, $\bS_{r,k}$ of dimension $\ns\times \ns$, and $\bR_d$ of dimension $\nr\times \nr$, as well as the transmit covariance matrices: $\bT_k$ of dimension $\nt\times \nt$, $\bS_{t,k}$ of dimension $\ns\times \ns$, and $\bT_d$ of dimension $\nt\times \nt$ are all non-negative definite having the following fixed traces: $\Tr{\bT_k}=\Tr{\bT_d}=\nt$, $\Tr{\bR_k}=\Tr{\bR_k}=\nr$, and $\Tr{\bS_{r,k}}=\Tr{\bS_{t,k}}=\ns$. Note that $\bS_{r,k}$ models the correlation of the incoming EM waves at the elements of each $k$-th RIS, while $\bS_{t,k}$ models the correlation at those elements for the outgoing (reflected) EM waves. For simplicity, we will not consider the polarization properties of the EM waves, treating them only as scalars. In the considered case, all above correlation matrices may be expressed in terms of a weight function $w(\bk)$ of the incoming or outgoing waves \cite{Moustakas2000_BLAST1_new}. $\bk$ is the corresponding $3$-dimensional wave vector with magnitude $|\bk|=k_0\triangleq\frac{2\pi}{\lambda}$, where $\lambda$ is the wavelength; $\bk$ can also be written as $\bk=[\bq,k_\perp]$, where $\bq$ is the $2$-dimensional component of $\bk$ along the RIS and $k_\perp$ is the amplitude perpendicular to it. Hence, we may also express $w(\bk)$ as $w(\bq,k_\perp)$. Following the latter notation, each $(a,b)$-element of $\bS_{r,k}$ $\forall k$ can be respectively obtained as:
\begin{equation}\label{eq:corr_mat_w(k)_def}
    \left[\bS_{r,k}\right]_{ab}=\int \, w_{r,k}(\bk) e^{i\bk^T(\bx_a-\bx_b)}d\Omega_{\bk},
\end{equation}
where $\bx_a$ and $\bx_b$ are the location coordinates of the respective elements of the $k$-th RIS. The above integral is taken over all directions of $\bk$ over the unit sphere with differential solid angle $d\Omega_\bk$, and $w_{r,k}(\bk)$ is normalized so that, when $\bx_a=\bx_b$, the integral gives unity. The other correlation matrices in \eqref{eq:all_correlations} can be expressed in a similar way. Note that the generic weight function $w(\bk)$ can be characterized by the mean direction of arrival or departure $\bs_0$ (with $|\bs_0|=k_0$), and the Angle Spread (AS) $\sigma$ (in radians), hence, we can write:
\begin{equation}\label{eq:weight_fn_def}
    w(\bk)\propto e^{-\frac{|\bk-\bs_0|^2}{2\sigma^2k_0^2}}.
\end{equation}

\section{Analytic Results}\label{sec:Mutual Information}\vspace{-0.1cm}
\subsection{Mutual Information (MI)}\vspace{-0.1cm}
The most relevant performance metric of communication links is the Mutual Information (MI), which can be expressed for our considered system model given by \eqref{eq:basic_channel_eq} as follows \cite{Foschini1998_BLAST1_mine}:
\begin{eqnarray}\label{eq:mut_info_def}
I \triangleq \log\det\left(\vec I_\nr + \rho\bG_{\rm tot} \bQ \bG_{\rm tot}^\dagger \right).
\end{eqnarray}
The above rate, expressed in nats per channel use, is achievable for Gaussian input signals with covariance matrix $\bQ$, assuming that the RX knows the channel matrix $\bG_{\rm tot}$ through pilot signaling \cite{wang2020channel}. Since the channel matrix fluctuates in time due to fading, its long-time performance is captured by its ergodic average, denoted by $\ex[I]$. One important result of this paper is the derivation of a closed-form expression for the ergodic MI, which is valid in the large $\nr$, $\nt$, and $\ns$ limit. The performance metric will generally depend on all $\bPhi_k$'s. Based on this fact, we will optimize those RIS reflection matrices in order to maximize the ergodic MI for the considered RISs-empowered communication system, subject to statistical knowledge of the channel, which is more reliable than the instantaneous one. The following proposition presents the asymptotic form of the MI's ergodic average.

\begin{Proposition}\label{prop:ergMI}
Let the channel matrix $\bG_{tot}$ be composed as in \eqref{eq:Gtot} with the $\bG_d$, $\bG_{r,k}$, and $\bG_{t,k}$ of the direct channel and the outgoing and incoming channels from each $k$-th RIS being complex Gaussian random matrices with covariance given by \eqref{eq:Gd_cov}, \eqref{eq:Grk_cov}, and \eqref{eq:Gtk_cov}, respectively. In the limit $\nt, \nr, \ns\to\infty$ with fixed ratios $\beta_r\triangleq\nr/\nt$ and $\beta_s\triangleq\ns/\nt$, the ergodic MI per RX antenna element takes the following form:
\begin{align}\label{eq:S0}
 &C \triangleq \frac{\ex[I]}{\nt}= 
  \frac{1}{\nt}\sum_{k=1}^K \log\det \left(\bI_{\ns} +  t_{1k}r_{2k}\gamma_k \bSigma_k  \right)\nonumber
  \\ 
  &\frac{1}{\nt}\log\det \left(\bI_\nr + \tilde{\bR}\right)+\frac{1}{\nt}\log\det \left(\bI_\nt + \rho\bQ\tilde{\bT}\right)
  \\ \nonumber
  &- r_d t_d-\sum_{k=1}^K\left(r_{1k}t_{1k}+r_{2k}t_{2k}\right),
\end{align}
where the matrices $\tilde{\bR}$, $\tilde{\bT}$, and $\bSigma_k$ are defined as: 
\begin{align}
    \tilde{\bR}&\triangleq r_d \bR_d + \sum_{k=1}^K r_{1k}\bR_k,\,\,
    \tilde{\bT}\triangleq t_d\bT_d+\sum_{k=1}^K t_{2k}\bT_k,\label{eq:R_T_tilde}\\
    \bSigma_k &\triangleq \bS_{t,k}^{1/2}\bPhi_k^\dagger\bS_{r,k}\bPhi_k\bS_{t,k}^{1/2}\label{eq:Sigma_k_initial},
\end{align} 
and the parameters $r_{1k}$, $t_{1k}$, $r_{2k}$, $t_{2k}$, $r_d$, and $t_d$ are the unique solutions of the following fixed-point equations:
\begin{equation}\label{eq:fp_eqs}
\begin{split}
    t_d &=\frac{1}{\nt}\Tr{(\bI_\nr+\tilde{\bR})^{-1}\bR_d},\\
    t_{1k} &=\frac{1}{\nt}\Tr{(\bI_\nr+\tilde{\bR})^{-1}\bR_k}, 
  \\ 
    r_d &= \frac{\rho}{\nt} \Tr{\bQ\bT_d\left(\bI_\nt + \rho\bQ\tilde{\bT}\right)^{-1}}, 
  \\ 
    r_{2k} &= \frac{\rho}{\nt} \Tr{\bQ\bT_{k}\left(\bI_\nt + \rho\bQ\tilde{\bT}\right)^{-1}}, 
  \\ 
   r_{1k}& = \frac{\gamma_kr_{2k}}{\nt} \Tr{\bSigma_k\left(\bI_{\ns} +
\gamma_kt_{1k}r_{2k}\bSigma_k \right)^{-1} }, 
  \\ 
   t_{2k} &= \frac{\gamma_kt_{1k}}{\nt} \Tr{\bSigma_k\left(\bI_{\ns} + \gamma_kt_{1k}r_{2k}\bSigma_k\right)^{-1} }.
\end{split}   
\end{equation}

\end{Proposition}
\begin{proof}
The proof is delegated in Appendix~\ref{app:proof_ergMI}. 
\end{proof}

\subsection{Optimization}
\begin{figure}[!t]
	\centering
	\includegraphics[width=0.44\textwidth]{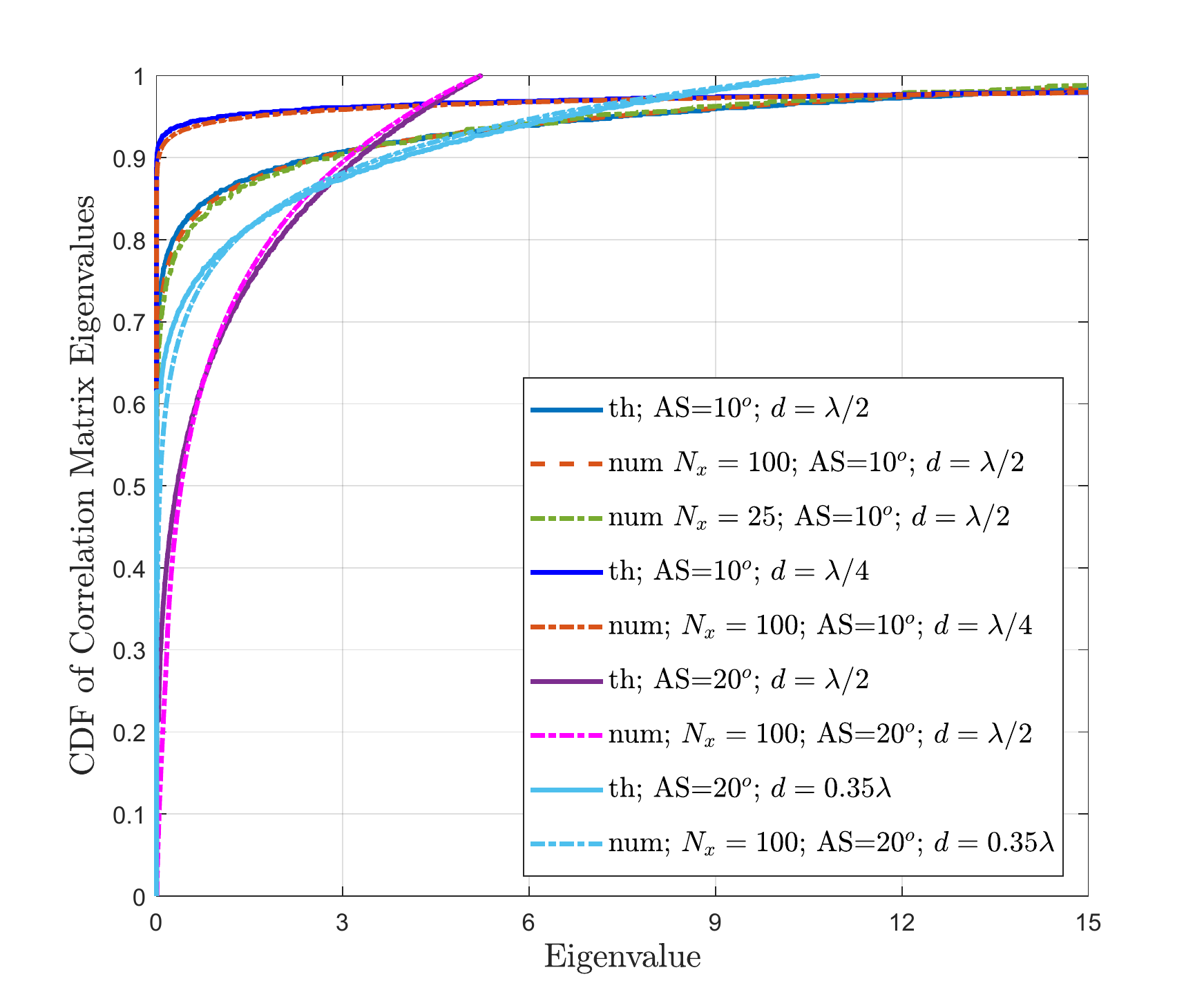}
	  \caption{Cumulative Distribution Function (CDF) of the eigenvalues of the correlation matrices $\bS_{t,k}$ and $\bS_{r,k}$. The theoretical curves are based on Proposition \ref{prop:eigenvalues} for the Gaussian weight function $w(\bk)$ in \eqref{eq:weight_fn_def} with $\mathbf{s}_0=k_0\hat{\be}_\perp$, where $\hat{\be}_\perp$ is the incoming wave direction vertical to the RIS. The agreement between the analytical (th) and numerical (num) results is very good.}\vspace{-0.6cm}
		\label{fig:cdf_corr_mat_eigs}
\end{figure}

\vspace{-0.1cm}
Having expressed the average MI in closed form in the previous proposition, our next task is to optimize it with respect to the elements of all $\bPhi_k$'s. Since each $\bPhi_k$ enters the calculation only through the matrix $\bSigma_k$, it is instructive to first understand the structure of this matrix, and more precisely, the spectrum of the correlation matrices $\bS_{t,k}$ and $\bS_{r,k}$ referring to each $k$-th RIS. We start by noting that each $\bSigma_k$ can be expressed in terms of the eigenvalues and eigenvectors of $\bS_{t,k}$ (denoted as $\{\eta_{n,k}\}_{n=1}^\ns$ and $\{\bu_{n,k}\}_{n=1}^\ns$) and $\bS_{r,k}$ (denoted as $\{\kappa_{n,k}\}_{n=1}^\ns$ and $\{\bv_{n,k}\}_{n=1}^\ns$) as follows:
\begin{align}
\label{eq:Sigma_k}
\bSigma_k&=\sum_{\ell,\ell'=1}^\ns \bv_{\ell,k}\bv^\dagger_{\ell',k} \sqrt{\kappa_{\ell,k}\kappa_{\ell',k}} \sum_{m=1}^\ns\eta_m \alpha_{m\ell , k} \alpha_{m\ell',k}^* ,
\nonumber \\
\alpha_{m\ell,k}&\triangleq\bu_{m,k}^\dagger\bPhi_k\bv_{\ell,k}.
\end{align}
The above expressions highlight that the effect of each $\bPhi_k$ on the MI is filtered through the eigenvectors of $\bS_{t,k}$ and $\bS_{r,k}$ in the form expressed by $\alpha_{m\ell,k}$. Hence, it will be useful to characterize them. In \cite{Moustakas2000_BLAST1_new}, it was first suggested that those eigenvectors are Fourier modes, and later also discussed in \cite{Pizzo2020_DOFHolographicMIMO}. In the following proposition, we make this claim more concrete, adding a technical constraint on the weight function $w(\bq,k_\perp)$ that vanishes for incoming (or outgoing) radiation along the RIS. This condition is met when the AS $\sigma$ is relatively small and does not qualitatively alter our model.
\begin{Proposition}\label{prop:eigenvalues}
Let the reflecting elements of an RIS form a square grid of length $L\triangleq\sqrt{\ns}\alpha$ with distance $\alpha\leq \lambda/2$ between nearest neighboring elements. Assume that the weight function $w(\bk)$ in \eqref{eq:corr_mat_w(k)_def} is such that $\lim_{|k_\perp|\to 0}w(\bq,k_\perp)/|k_\perp|<\infty$. Then, in the limit $\ns\to\infty$, the eigenvectors of the incoming and outgoing covariance matrices are Fourier modes, i.e., $[\bu_{n}]_\ell\propto e^{i\bq_n^T\bx_\ell}$ for $n,\ell=1,2,\ldots,\ns$, where $\bq_n\triangleq2\pi[m_1\,m_2]/L$ (with integers $|m_1|,|m_2|<L/(2a)$) is a Fourier vector on the RIS plane. The corresponding eigenvalue, parameterized by $\bq_n$, takes the following value: 
\begin{equation}\label{eq:prop:eigenvalue}
\eta_n\triangleq \left(\frac{\lambda}{a}\right)^2\frac{w(\bq_n,k_\perp)+w(\bq_n,-k_\perp)}{\sqrt{1-\frac{|\bq_n|^2}{k_0^2}}} \Theta(k_0-|\bq_n|),
\end{equation}
with $k_\perp=\sqrt{k_0^2-|\bq_n|^2}$ where $\Theta(x)$ is the step function, with $\Theta(x)=1$ if $x>0$ and zero if $x<0$. 
\end{Proposition}
\begin{proof}
The assumption that $w(\bq,k_\perp)$ vanishes when $|\bq|\to k_0$ ensures that the Dirichlet-Dini criterion holds for $\eta_n$ in \eqref{eq:prop:eigenvalue}. In this case, the Fourier transform of both correlation matrices converges to $\eta_n$. For small ASs and angles of arrival away from the horizontal, this assumption is immaterial. 
\end{proof}
Since the weight function $w(\bk)$ is negligible outside the region $|\bk-\bs_0|\leq \frac{k_0\sigma}{2\pi}$, we may estimate the number of non-negligible eigenvalues to be roughly $\frac{\sigma^2 L^2}{(2\pi)^2\lambda^2}$. In Fig. \ref{fig:cdf_corr_mat_eigs}, we plot the distribution function of the eigenvalues for various ASs and RIS element spacings. We see very good agreement of the above theoretical formula for the eigenvalues with the numerical diagonalization of the correlation matrices. Below, we will find convenient to order the eigenvalues from largest to the smallest, together with the corresponding eigenvectors.   

Using Proposition~2, we can analyze two limiting cases for the AS of the correlation matrices. In the first case, the AS at each $k$-th RIS is very high, making $\bS_{t,k}$ and $\bS_{r,k}$ essentially proportional to the unit matrix. In this case, the optimization over $\bPhi_k$'s is immaterial, since $\bSigma_k=\bPhi_k^\dagger\bPhi_k=\bI_\ns$. In the opposite case, $\bS_{t,k}$ and $\bS_{r,k}$ are unit-rank matrices, corresponding effectively to a Line-Of-Sight (LOS) channel. In this case, it holds $\forall a,b$ that $[\bS_{r,k}]_{a,b}=\ns[\bu_{1,k}\bu_{1,k}^\dagger]_{a,b}=e^{i\bq_{1r}^T(\bx_a-\bx_b)}$ and $[\bS_{t,k}]_{a,b}=\ns [\bv_{1,k}\bv_{1,k}^\dagger]_{a,b}=e^{i\bq_{1t}^T(\bx_a-\bx_b)}$, yielding the following expression for $\bSigma_k$ appearing in \eqref{eq:Sigma_k}:  
\begin{align}\label{eq:unit_rank_Sigma}
    \bSigma_k&= \ns |\alpha_{11,k}|^2 \bv_{1,k}\bv^\dagger_{1,k},
    \\ \nonumber 
    \alpha_{11,k}&=\bu_{1,k}^\dagger\bPhi_k\bv_{1,k}=\frac{1}{\ns}\sum_{n=1}^{\ns} e^{i\theta_{k,n}}e^{-i(\bq_{1t}-\bq_{1r})^T\bx_n}.
\end{align}

The latter expression implies that the $\bPhi_k$'s maximizing the ergodic MI are such that $\theta_{k,n}=(\bq_{1t}-\bq_{1r})^T\bx_n$ $\forall$$k,n$. We can clearly see that, in the case of geometric optics for which the components of the incoming and outgoing direction vectors parallel to the RIS are equal (i.e., $\bq_{1t}=\bq_{1r}$), no phase optimization is necessary. However, when this is not the case, optimization will produce significant gains. Interestingly, the Fourier form of the eigenvectors will allow us to gain insight on the optimization process and obtain the optimal $\bPhi_k$'s in closed form. Indeed, when the eigenvalue distributions of $\bS_{r,k}$ and $\bS_{t,k}$ are a displacement of one another in the $\bq$-space, as seen in the insert figure of Fig.~\ref{fig:MI_AS_RISsizes}, then the differences of the $\bq$-vectors of the corresponding eigenvalues in each distribution are constant and equal to the difference between the $\bq$-vectors of the maximum eigenvalues of the matrices, i.e., $\bq_{\ell t}-\bq_{\ell r}=\bq_{1t}-\bq_{1r}$ $\forall$$\ell\neq1$. Thus, setting the phases of all $\bPhi_k$'s for the $K$ RISs as follows:
\begin{equation}\label{eq:Phi_opt}
e^{i\theta_{k,n}}=e^{i(\bq_{1t}-\bq_{1r})^T\bx_n}\,\,\forall k,n,   
\end{equation}
resulting using \eqref{eq:Sigma_k} in $\alpha_{m\ell,k}=\delta_{m,\ell}$ ($\delta_{m,\ell}$ is the Kronecker delta function) $\forall$$m,\ell=1,2,\ldots,N_s$,
is optimal. Clearly, if the distribution of the eigenvalues is not exactly the same, then the above conjecture for the RIS elements' values may not be optimal, but will be close to them, and thus, can serve as initial conditions for further optimization.

\section{Numerical Results}\label{sec:Optimization of RISs}\vspace{-0.1cm}
We now wish to test our previous analytical results versus brute-force optimization of the RISs. To this end, we use the closed-form expression \eqref{eq:S0} for the ergodic MI to design optimal $\bPhi_k$'s for all $K$ RISs that maximize it. By expressing this performance metric as a function of all RIS reflection matrices, we aim at solving the following Optimization Problem (OP) where $n = 1,2,\dots,\ns$:
\begin{align*} 
\begin{split}
    \mathcal{OP}_1: \,\, \max_{\{\bPhi_k\}_{k=1}^K} \,\,C\left(\{\bPhi_k\}_{k=1}^K \right) \quad\text{s.t.} \quad \lvert [\bPhi_k]_{n,n} \rvert = 1  \, \, \forall k,n.
\end{split}
\end{align*}
For this goal, we adopt an iterative approach based on alternating optimization, as follows. First, it is noted that each $\bPhi_k$ is included in the expression \eqref{eq:Sigma_k_initial} for $\bSigma_k$, as well as in the expressions \eqref{eq:R_T_tilde} for $r_{1k}$, $t_{1k}$, $r_{2k}$, $t_{2k}$, $r_d$, and $t_d$. To this end, in each algorithmic iteration, we first consider that the latter variables are fixed and solve for $\bPhi_k$'s maximizing MI. By inspecting \eqref{eq:S0}, this problem simplifies as follows:  
\begin{align*} 
\begin{split}
    \mathcal{OP}_2: \,\, \max_{\{\bPhi_k\}_{k=1}^K} &\,\,\sum_{k=1}^K \log\det \left(\bI_{\ns} +  
t_{1k}r_{2k}\bPhi_k^\dagger\bS_{r,k}\bPhi_k\bS_{t,k}  \right) \\
    \hspace{0.4cm}\text{s.t.} \quad & \lvert [\bPhi_k]_{n,n} \rvert = 1  \, \, \forall k,n,
\end{split}
\end{align*}
which can be solved via \cite{Zhang_Capacity}; the detailed solution and algorithm are omitted here due to space limitations. Afterwards, we substitute the $\mathcal{OP}_2$ solution into \eqref{eq:fp_eqs} to calculate $r_{1k}$, $t_{1k}$, $r_{2k}$, $t_{2k}$, $r_d$, and $t_d$. The latter two steps are repeated at each iteration until convergence, or when reaching a threshold value for the ergodic MI objective.

\begin{figure}[!t]
	\centering
	\includegraphics[width=0.5\textwidth]{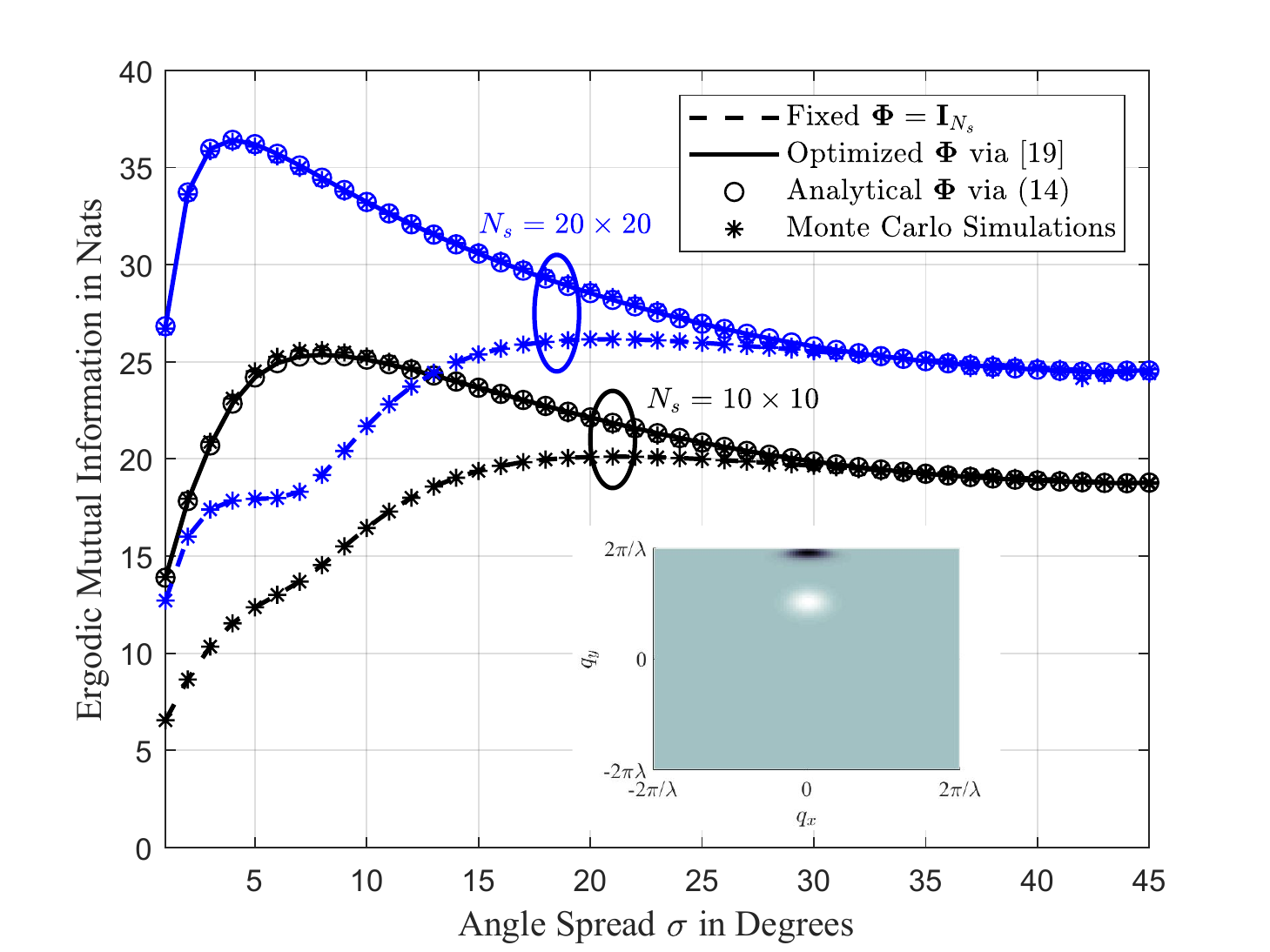}
	  \caption{The ergodic mutual information in nats per channel use for $\nt=8$, $\nr=4$, and $\rho=10$ dB versus the AS in degrees, considering one RIS operating at $2.5$ GHz with reflecting elements of inter-element distance $\lambda/2=6$ cm and the case where there is no direct link between TX and RX. RIS optimization plays a prominent role in low AS values and large $\ns$ values (i.e., correlated channels), while for large ASs, the RIS optimization becomes unnecessary. The inset depicts the distribution of the eigenvalues of the incoming waves (white) and outgoing waves (black) with angles $30^o$ and $70^o$, respectively, and angles spread $\sigma=5^o$.}\vspace{-0.4cm}
		\label{fig:MI_AS_RISsizes}
\end{figure}

\begin{figure}[!t]
	\centering
	\includegraphics[width=0.5\textwidth]{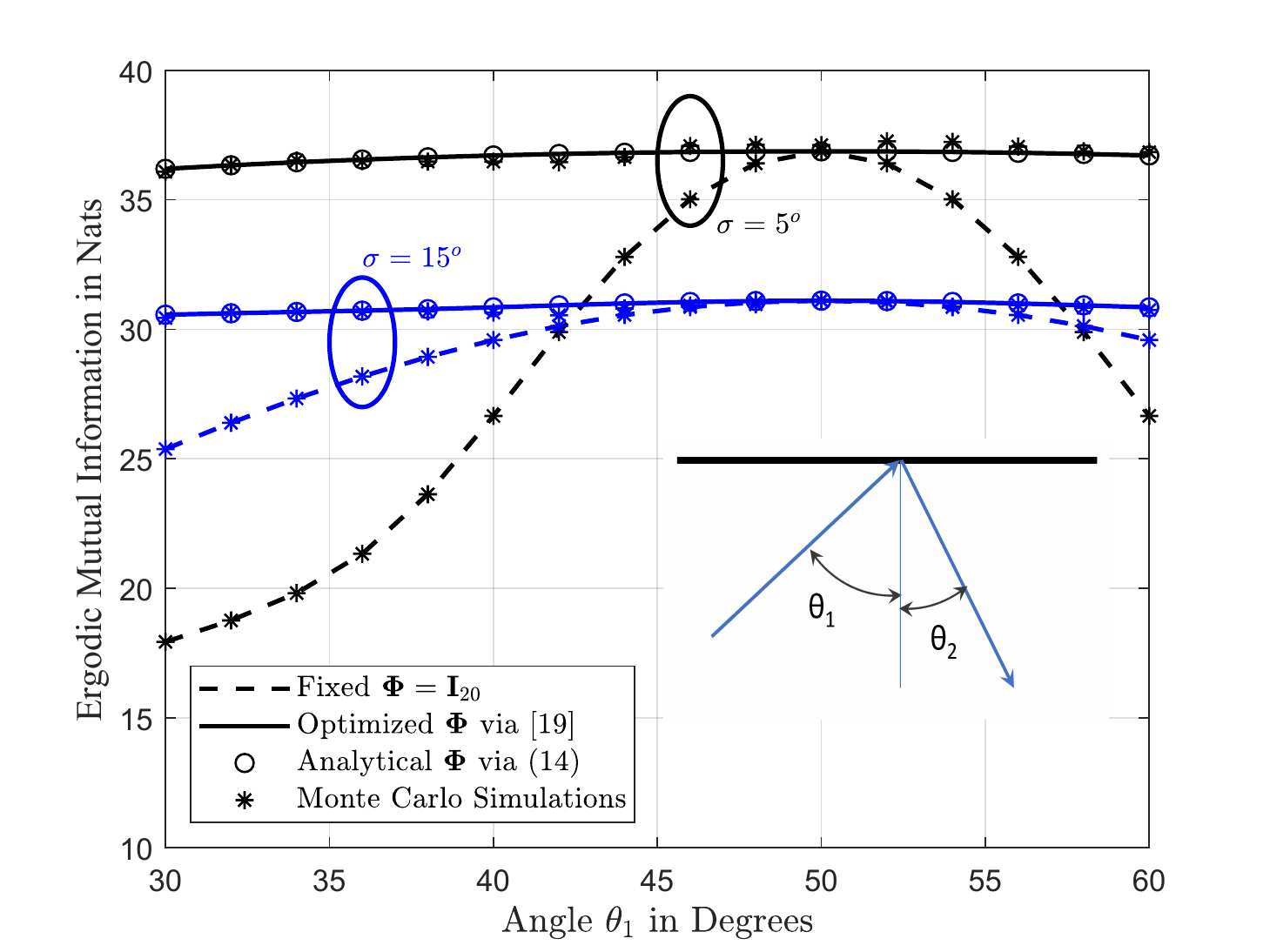}
	  \caption{The ergodic mutual information in nats per channel use versus the incoming wave's angle $\theta_1$ at the $20\times20$ RIS for SNR $\rho=10$ dB, considering a low and a moderate values for AS. The angle between the incoming and outgoing waves at the RIS was set to $100^o$, and the angle of the outgoing wave is $\theta_2=100-\theta_1$. The figure shows that, in the low AS case for fixed $\mathbf{\Phi}$, the maximum capacity is achieved when $\theta_1=\theta_2$ (geometrical optics), which is also obtained by the proposed optimization.}\vspace{-0.4cm}
		\label{fig:MI_theta1_ASs}
\end{figure}

To assess the effect of the proposed RIS optimization, we calculate the ergodic MI in the presence of one RIS. For simplicity, we do not include a direct path $\bG_d$ and assume that the TX and RX arrays are uncorrelated. In Fig$.$~\ref{fig:MI_AS_RISsizes}, we plot the MI as a function of the AS $\sigma$ when geometric optics is not possible, as well as before and after optimization over $\bPhi$. We see that for ASs up to about $20^o$, optimization over $\bPhi$ provides significant gains, irrespective of the size of the RIS. In fact, for ASs around $5^o-10^o$, the ergodic MI is higher than the one with no correlations. This surprising phenomenon is due to the significant ``beamforming'' gain from the coherent reflection from the RIS along the directions of the non-negligible eigenvalues. 

Furthermore, we explore another important aspect of the RIS optimization. As seen in the previous section and \eqref{eq:unit_rank_Sigma}, when the correlation matrices $\bS_{t,k}$ and $\bS_{r,k}$ are unit-rank, the degree to which optimization plays a role depends on whether the desired impinging and outgoing directions are those corresponding to geometrical optics. In Fig. \ref{fig:MI_theta1_ASs}, we plot the ergodic MI as a function of the incoming angle $\theta_1$, with respect to the vertical of the single RIS, for fixed desired angular difference between the incoming and outgoing mean directions (corresponding to the sum $\theta_1+\theta_2$ in the figure). Note that, varying $\theta_1$ is equivalent to rotating the RIS around a horizontal axis. We see that, when $\theta_1=\theta_2$, which corresponds to geometrical optics, optimization gives no gains. The gains can be significant though for other RIS  orientations. Interestingly, in both above figures, brute-force optimization using \cite{Zhang_Capacity} produces identical results with the setting $\theta_{n}=\angle([\bu_{1}]_n[\bv_{1}^*]_n)$ for each $n$-th RIS element with $n=1,2,\ldots,\ns$. This corresponds to the phase difference between the elements of the maximum-eigenvalue eigenvectors, as discussed in the previous section. 

\section{Conclusion}
\label{sec:conclusion}\vspace{-0.15cm}
In this paper, we have presented a novel expression for the asymptotic ergodic MI performance of multi-antenna wireless communication systems in the presence of multiple RISs in the limit of large antenna arrays, using random matrix theory and methods from statistical physics. When fading is present, it makes more sense to optimize the phases of the RIS reflecting elements based on the statistical knowledge of the channel. In this case, we have shown that finite AS can play a crucial role in the degree to which the RISs can be optimized. While for large ASs optimization is unnecessary, for low AS values, which is reasonable for increasing carrier frequencies which exhibit reduced multipath, significant capacity gains can be obtained with RIS optimization. Correspondingly, phase optimization plays an important role when the required incoming and outgoing directions of the signal at the RIS are significantly different from the ones prescribed by geometrical optics. Moving forward, it is important to analyze the effect on capacity, interference, and the limitations of RISs' optimization in the presence of multiple TX-RX arrays, sharing the wireless medium and the RISs, which lies at the heart of the success of RISs for 6G communications. The model described here can be readily applied to assess the impact of multiple RISs in range extension, propagation around obstacles, as well as the limitations of the use of RISs with multiple TX-RX pairs. 
\vspace{-0.1cm}

\appendix
\section{Proof of Proposition \ref{prop:ergMI}}
\label{app:proof_ergMI}\vspace{-0.2cm}

To prove this proposition we will use methods from random matrix theory and the replica approach, a technique originally developed in
the context of statistical physics and successfully applied to several problems in wireless communications \cite{Tanaka2002_ReplicasInCDMAMUD, Moustakas2003_MIMO1_mine, Guo2003_ReplicaAnalysisOfLargeCDMASystems, Muller2003_RandomMatrixMIMO_binary_mine, Taricco2008_MIMOCorrelatedCapacity}. Since the method has already been analyzed in the past, we will only provide highlights of the proof. We first define the following scalar quantity:
\begin{equation}
    {\cal Z}=\det\left(\bI_\nr+\rho\bG_{\rm tot}\bQ\bG_{\rm tot}^\dagger\right)^{-1},
\end{equation}
using the corresponding letter (${\cal Z}$), which is used in statistical physics for the so-called partition function of the system. Then, the moment generating function of the MI can be expressed as $g(\nu)=\ex\left[{\cal Z}^\nu\right]$. From this expression, the asymptotic normalized MI can be expressed as
\begin{equation}\label{eq:c=loggnu}
    C=-\lim_{\nt\to\infty} \nt^{-1}\left.\left( \log g(\nu)\right)'\right|_{\nu=0^+}.
\end{equation}
To make progress, we will make a number of assumptions, which, while not proven, have been conjectured to be valid.
\begin{assumption}
The calculation of $g(\nu)$ for $\nu\in\ZZ^+$ can be analytically continued to real values of $\nu\in(0,1)$.
\end{assumption}
\begin{remark*}\rm
This property will allow us to obtain the behavior close to $\nu=0^+$ by evaluating the expression for integer values of the replica index $\nu$.
\end{remark*}
We will now start with the evaluation of $g(\nu)$ for integer $\nu\in\ZZ^+$. It is straightforward to show the following expression (see \cite{Moustakas2007_MIMO1} for details):
\begin{align}
    &{\cal Z}^\nu = \ex\left[
    e^{\frac{1}{2}\Tr{\bY^\dagger\bG_{d}^\dagger  \bZ-\bZ^\dagger \bG_{d}\bQ\bY }}\times\right. 
\\ \nonumber
        &\left.\!
         e^{\frac{1}{2}\!\!\sum\limits_{k=1}^K\!\!\Tr{\bV_k^\dagger\bG_{t,k}\bY-\bY^\dagger \bQ\bG_{t,k}^\dagger\bW_k-\bZ^\dagger\bG_{r,k}\bPhi_k\bV_k-\bW_k^\dagger\bPhi^\dagger_k\bG_{r,k}^\dagger\bZ}}
        \right]\!\!,
\end{align}
where the expectation is over the complex Gaussian matrices $\bZ$, $\bY$, $\bV_k$, and $\bW_k$ of dimensions $\nr\times \nu$, $\nt\times\nu$, $\ns\times\nu$ and $\ns\times\nu$, respectively, with variance equal to $2$. In the above expression, the channel matrices appear in the exponent of the exponential in a linear fashion, and hence, they may be integrated out resulting to:
\begin{align}
\label{eq:gnu1}
&g(\nu) =  \ex\left[ e^{ -\frac{1}{4\nt}\Tr{\bY^\dagger\bQ\bT_d\bY\bZ^\dagger\bR_d\bZ} } 
\right.
\\ \nonumber
&\left.\!\times e^{\frac{1}{4\nt} \sum_{k=1}^K\Tr{\bZ^\dagger\bR_k\bZ\bW_k^\dagger\bPhi_k^\dagger\bS_{rk}\bPhi_k\bV_k - \bY^\dagger\bQ\bT_k\bY\bV_k^\dagger\bS_{tk}\bW_k}} \right]\!\!.
\end{align}
The integration over the Gaussian channel matrices has resulted into terms in the exponent, which are quartic in the Gaussian random variables. To overcome this difficulty we employ the following identity, which decomposes them into quadratic terms via the Fourier representation of the Dirac delta-function.
\begin{identity}
\label{id:hub_strat}%
If $\bA, \bB\in\CC^{\nu\times\nu}$ the following identity holds \cite{Taricco2008_MIMOCorrelatedCapacity}:
\begin{align}
\label{eq:hub_strat_identity}%
&e^{-\frac{1}{\nt}\Tr{\bA\bB}}  =
\\ \nonumber
& \lim_{\epsilon\to 0^+}\int e^{ \Tr{\nt\left(-\epsilon{\mathbfcal{T}  \mathbfcal{T}}^T+\epsilon{\mathbfcal{R} \mathbfcal{R}}^T+{\mathbfcal{R} \mathbfcal{T}}\right)-  \bA {\mathbfcal  T} - {\mathbfcal R}
\bB }}d\mu({\mathbfcal  T}, {\mathbfcal
R}),
\end{align}
where the integration metric $d\mu({\mathbfcal T, \bRcal})$ is given by
\begin{equation}
d\mu({\mathbfcal{T}, \mathbfcal{R}}) = \prod_{\alpha,\beta=1}^\nu \nt 
\frac{d{\mathbfcal T}_{\alpha\beta} d{\mathbfcal R}_{\beta\alpha}}{2\pi i}    
\end{equation}
and the integration of the elements of ${\mathbfcal T}$ is over the real axis, while the integration of the ${\mathbfcal R}$ is over the imaginary axis.
\end{identity}

Following a standard procedure \cite{Moustakas2003_MIMO1_mine}, we introduce the matrices: \textit{i}) ${\mathbfcal T}_d$ and ${\mathbfcal R}_d$ to decompose the term in the first line of \eqref{eq:gnu1} with $\bA=\frac{1}{2\nt}\bZ^\dagger\bR_d\bZ$ and $\bB=\frac{1}{2}\bY^\dagger\bQ\bT_d\bY$; \textit{ii}) the matrices ${\mathbfcal T}_{1k}$ and ${\mathbfcal R}_{1k}$, with $k=1,2,\ldots,K$, to decompose the first term in the second line of \eqref{eq:gnu1} with $\bA=-\frac{1}{2}\bW_k^\dagger\bPhi_k^\dagger\bS_{rk}\bPhi_k\bV_k$ and $\bB=\frac{1}{2\nt}\bZ^\dagger\bR_k\bZ$; and finally \textit{iii}) the matrices ${\mathbfcal T}_{2k}$ and ${\mathbfcal R}_{2k}$, with $k=1,2,\ldots,K$, to decompose the second term in the second line of \eqref{eq:gnu1} with $\bA=\frac{1}{2\nt}\bY^\dagger\bQ\bT_k\bY$ and $\bB=\frac{1}{2}\bV_k^\dagger\bS_{tk}\bV_k$. 
This allows us to integrate over the complex matrices $\bZ$, $\bY$, $\bV_k$, and $\bW_k$. As a result, \eqref{eq:gnu1} becomes:
\begin{align}
\label{eq:gnu2} %
g(\nu)=\int e^{-{\cal S }}d\mu(\{\bTcal, \bRcal\})
\end{align}
with $d\mu(\{\bTcal, \bRcal\})$ indicating an integration over all ${
\mathbfcal{T}, \bRcal}$ matrices introduced above, and where the exponent ${\cal S}$ takes the following form:
\begin{align}
& {\cal S} = \log\det \left(\bI_\nr \outer \bI_{\nu} + \left(\bR_d \outer \bRcal_d+
  \sum_k \bR_k \outer \bRcal_{1k} \right)\right)
\nonumber \\
&+\log\det \left(\bI_\nt \outer \bI_{\nu} + \left(\bQ\bT_d \outer \bTcal_d+
  \sum_k \bQ\bT_k \outer \bTcal_{2k} \right)\right)
\nonumber \\
 &-\sum_k \log\det \left(\bI_\nu \outer \bI_{\ns} +  
\bTcal_{1k}\bRcal_{2k}\outer \bPhi_k\bS_{rk}\bPhi_k\bS_{tk}  \right)
\nonumber \\%
 &- \nt\Tr{\bTcal_d \bRcal_d+ \sum_k \left(\bTcal_{1k} \bRcal_{1k} + \bTcal_{2k} \bRcal_{2k}\right) },
\label{eq:S_full}
\end{align}
where the notation $\outer$ denotes the direct product of matrices. Since the integral in \eqref{eq:gnu2} cannot be performed exactly, we will evaluate it asymptotically for large antenna numbers.
\begin{assumption}
The analytic continuation of $g(\nu)$ to real $\nu$ values in \eqref{eq:c=loggnu} can be interchanged with the limit $\nt\to\infty$.
\end{assumption}
To obtain the asymptotic evaluation of \eqref{eq:gnu2}, we will deform the contours of the integrals of the elements of $\{\mathbfcal{T}, \mathbfcal{ R}\}$ to pass through a saddle point of ${\cal S}$. More details can be found in \cite{Moustakas2003_MIMO1_mine, Bender_Orszag_book}. To proceed we need to specify the structure of the saddle-point and the symmetry of the values of the matrices there. Since the $\nu$ replicas are introduced \textit{a priori} equivalent with each other, it is natural to make the assumption that follows. 
\begin{assumption}
At the saddle point, the matrices $\{\mathbfcal{T}, \mathbfcal{ R}\}$ appearing in \eqref{eq:S_full} are rotationally invariant under continuous replica rotations and are thus proportional to $\bI_\nu$.
\end{assumption}
\begin{remark*}
\rm This technical assumption, which seems intuitively obvious, very often becomes invalid in certain systems, in which this symmetric solution becomes unstable, leading to the so-called replica-symmetry breaking phenomenon \cite{ParisiBook}. Nevertheless, in this case, due to the continuous symmetry present, the symmetry can be shown to be always stable \cite{Moustakas2007_MIMO1}, giving extra credence to the result.
\end{remark*}
Given the above assumption, we have:
\begin{align}
    \left.{\mathbfcal T}_d\right|_{saddle point}=t_d\bI_\nu &\mbox{,}\,\,\,\,\,\,\,\,\,\, \left.{\mathbfcal R}_d\right|_{saddle point}=r_d\bI_\nu, 
    \\ \nonumber
    \left.{\mathbfcal T}_{ak}\right|_{saddle point}=t_{ak}\bI_\nu &\mbox{,}\,\,\,\,\,\,\,\,\,\, \left.{\mathbfcal R}_{ak}\right|_{saddle point}=r_{ak}\bI_\nu, 
\end{align}
where $a=1,2$ and $k=1,2,\ldots,K$. To obtain the values of $r_d$, $t_d$, $r_{ak}$, and $t_{ak}$, we need to solve the saddle-point equations, which can be derived by demanding that ${\cal S}$ is stationary with respect to variations of $\{{\mathbfcal T, \bRcal}\}$ (see \cite{Bender_Orszag_book} for details). This produces the fixed-point equations in \eqref{eq:fp_eqs}, which can be shown to have unique solutions \cite{Taricco2008_MIMOCorrelatedCapacity}. Hence, $g(\nu)$ can be written as follows:
\begin{equation}
    g(\nu)=e^{-{\cal S}_0} \int e^{-({\cal S}-{\cal S}_0)} d\mu(\{{\mathbfcal{T}, \mathbfcal{R}}\}),
\end{equation}
where ${\cal S}_0\triangleq\nu \nt C$ is the saddle-point value of ${\cal S}$ and $C$ is given in \eqref{eq:S0}. The remaining integral can be shown to give $O(\nt^{-1})$ corrections to the ergodic average of the MI \cite{Moustakas2003_MIMO1_mine, Moustakas2007_MIMO1} and can also provide its variance, by integrating over the second-order corrections of $\{{\mathbfcal{T}, \mathbfcal{R}}\}$ around the saddle point. We will leave this analysis for a future work. 

\scriptsize{
\section*{Acknowledgment}
This work has been partially supported by the EU H2020 RISE-6G project under grant number 101017011.}\vspace{-0.12cm}

\bibliographystyle{IEEEtran}\vspace{-0.1cm}
\bibliography{IEEEfull,wireless,references}

\begin{thebibliography}{10}
\providecommand{\url}[1]{#1}
\csname url@samestyle\endcsname
\providecommand{\newblock}{\relax}
\providecommand{\bibinfo}[2]{#2}
\providecommand{\BIBentrySTDinterwordspacing}{\spaceskip=0pt\relax}
\providecommand{\BIBentryALTinterwordstretchfactor}{4}
\providecommand{\BIBentryALTinterwordspacing}{\spaceskip=\fontdimen2\font plus
\BIBentryALTinterwordstretchfactor\fontdimen3\font minus
  \fontdimen4\font\relax}
\providecommand{\BIBforeignlanguage}[2]{{%
\expandafter\ifx\csname l@#1\endcsname\relax
\typeout{** WARNING: IEEEtran.bst: No hyphenation pattern has been}%
\typeout{** loaded for the language `#1'. Using the pattern for}%
\typeout{** the default language instead.}%
\else
\language=\csname l@#1\endcsname
\fi
#2}}
\providecommand{\BIBdecl}{\relax}
\BIBdecl
\renewcommand{\BIBentryALTinterwordstretchfactor}{4}

\bibitem{Samsung}
``The next hyper- {C}onnected experience for all,'' White Paper, Samsung 6G
  Vision, Jun. 2020.

\bibitem{huang2019holographic}
C.~Huang \emph{et~al.}, ``Holographic {MIMO} surfaces for 6{G} wireless
  networks: {O}pportunities, challenges, and trends,'' \emph{IEEE Wireless
  Commun.}, vol.~27, no.~5, pp. 118--125, Oct. 2021.

\bibitem{DMA_2020}
N.~Shlezinger \emph{et~al.}, ``Dynamic metasurface antennas for {6G} extreme
  massive {MIMO} communications,'' \emph{IEEE Wireless Commun.}, 2021.

\bibitem{liaskos2018new}
C.~Liaskos \emph{et~al.}, ``A new wireless communication paradigm through
  software-controlled metasurfaces,'' \emph{IEEE Commun. Mag.}, vol.~56, no.~9,
  pp. 162--169, Sep. 2018.

\bibitem{huang2019reconfigurable}
C.~Huang \emph{et~al.}, ``Reconfigurable intelligent surfaces for energy
  efficiency in wireless communication,'' \emph{IEEE Trans. Wireless Commun.},
  vol.~18, no.~8, pp. 4157--4170, Aug. 2019.

\bibitem{RIS_Scattering}
G.~C. Alexandropoulos \emph{et~al.}, ``Reconfigurable intelligent surfaces for
  rich scattering wireless communications: {R}ecent experiments, challenges,
  and opportunities,'' \emph{IEEE Commun. Mag.}, to appear, 2021.

\bibitem{WavePropTCCN}
G.~C. Alexandropoulos, G.~Lerosey \emph{et~al.}, ``Reconfigurable intelligent
  surfaces and metamaterials: {T}he potential of wave propagation control for
  {6G} wireless communications,'' \emph{IEEE ComSoc TCCN Newslett.}, vol.~6,
  no.~1, pp. 25--37, Jun. 2020.

\bibitem{smith2}
D.~R. Smith, J.~B. Pendry \emph{et~al.}, ``Metamaterials and negative
  refractive index,'' \emph{Science}, vol. 305, no. 5685, pp. 788--792, 2004.

\bibitem{smith3}
D.~R. Smith \emph{et~al.}, ``Analysis of a waveguide-fed metasurface antenna,''
  \emph{Physical Review Applied}, vol.~8, no.~5, 2017.

\bibitem{Hum2014}
S.~V. Hum and J.~Perruisseau-Carrier, ``Reconfigurable reflectarrays and array
  lenses for dynamic antenna beam control: {A} review,'' \emph{IEEE Trans.
  Antennas Prop.}, vol.~62, no.~1, pp. 183--198, Jan. 2014.

\bibitem{Science_2011}
N.~Yu \emph{et~al.}, ``Light propagation with phase discontinuities:
  {G}eneralized laws of reflection and refraction,'' \emph{Science}, vol. 334,
  no. 6054, pp. 333--337, 2011.

\bibitem{Jung2020}
M.~Jung \emph{et~al.}, ``Performance analysis of large intelligent surfaces
  ({LISs}): {A}symptotic data rate and channel hardening effects,'' \emph{IEEE
  Trans. Wireless Commun.}, vol.~19, no.~3, pp. 2052--2065, Mar. 2020.

\bibitem{Nadeem2020}
Q.~Nadeem \emph{et~al.}, ``Asymptotic max-min {SINR} analysis of reconfigurable
  intelligent surface assisted {MISO} systems,'' \emph{IEEE Trans. Wireless
  Commun.}, vol.~19, no.~12, pp. 7748--7764, Dec. 2020.

\bibitem{Selimis2021}
D.~Selimis \emph{et~al.}, ``On the performance analysis of {RIS}-empowered
  communications over {N}akagami-$m$ fading,'' \emph{IEEE Commun. Let.}, 2021.

\bibitem{wang2020channel}
Z.~Wang, L.~Liu, and S.~Cui, ``Channel estimation for intelligent reflecting
  surface assisted multiuser communications: Framework, algorithms, and
  analysis,'' \emph{{IEEE} Trans. Wireless Commun.}, vol.~19, no.~10, pp.
  6607--6620, Oct. 2020.

\bibitem{Moustakas2000_BLAST1_new}
A.~L. Moustakas \emph{et~al.}, ``Communication through a diffusive medium:
  {C}oherence and capacity,'' \emph{Science}, vol. 287, pp. 287--290, Jan.
  2000.

\bibitem{Foschini1998_BLAST1_mine}
G.~J. Foschini and M.~J. Gans, ``On limits of wireless communications in a
  fading environment when using multiple antennas,'' \emph{Wireless Personal
  Commun.}, vol.~6, pp. 311--335, 1998.

\bibitem{Pizzo2020_DOFHolographicMIMO}
A.~Pizzo \emph{et~al.}, ``Degrees of freedom of holographic {MIMO} channels,''
  in \emph{Proc. IEEE SPAWC}, Atlanta, USA, May 2020.

\bibitem{Zhang_Capacity}
S.~Zhang and R.~Zhang, ``Capacity characterization for intelligent reflecting
  surface aided {MIMO} communication,'' \emph{IEEE J. Sel. Areas Commun.},
  vol.~38, no.~8, pp. 1823--1838, Aug. 2020.

\bibitem{Tanaka2002_ReplicasInCDMAMUD}
T.~Tanaka, ``A statistical-mechanics approach to large-system analysis of
  \mbox{CDMA} multiuser detectors,'' \emph{{IEEE} Trans. Inform. Theory},
  vol.~48, no.~11, pp. 2888--2910, Nov. 2002.

\bibitem{Moustakas2003_MIMO1_mine}
A.~L. Moustakas \emph{et~al.}, ``\mbox{MIMO} capacity through correlated
  channels in the presence of correlated interferers and noise: \mbox{A} (not
  so) large \mbox{N} analysis,'' \emph{{IEEE} Trans. Inform. Theory}, vol.~49,
  no.~10, pp. 2545--2561, Oct 2003.

\bibitem{Guo2003_ReplicaAnalysisOfLargeCDMASystems}
D.~Guo and S.~Verd{\'u}, ``Randomly spread {CDMA}: {A}symptotics via
  statistical physics,'' \emph{{IEEE} Trans. Inform. Theory}, vol.~51, pp.
  1982--2010, Jun. 2005.

\bibitem{Muller2003_RandomMatrixMIMO_binary_mine}
R.~R. M\"uller, ``Channel capacity and minimum probability of error in large
  dual antenna array systems with binary modulation,'' \emph{IEEE Trans. on
  Signal Processs.}, vol.~51, no.~11, pp. 2821--2828, Nov. 2003.

\bibitem{Taricco2008_MIMOCorrelatedCapacity}
G.~Taricco, ``Asymptotic mutual information statistics of separately-correlated
  {MIMO} {R}ician fading channels,'' \emph{{IEEE} Trans. Inform. Theory},
  vol.~54, no.~8, p. 3490, Aug. 2008.

\bibitem{Moustakas2007_MIMO1}
A.~L. Moustakas and S.~H. Simon, ``On the outage capacity of correlated
  multiple-path {MIMO} channels,'' \emph{{IEEE} Trans. Inform. Theory},
  vol.~53, no.~11, pp. 3887--3903, Nov. 2007.

\bibitem{Bender_Orszag_book}
C.~M. Bender and S.~A. Orszag, \emph{Advanced Mathematical Methods for
  Scientists and Engineers}.\hskip 1em plus 0.5em minus 0.4em\relax New York,
  NY: McGraw-Hill, 1978.

\bibitem{ParisiBook}
M.~M\'ezard, G.~Parisi, and M.~A. Virasoro, \emph{Spin Glass Theory and
  Beyond}.\hskip 1em plus 0.5em minus 0.4em\relax Singapore: World Scientific,
  1987.

\end{thebibliography}

\end{document}